# Inverse-design topology optimization of magnonic devices using level-set method

Check for updates

Andrey A. Voronov[1,2] ✉, Marcos Cuervo Santos[2,3], Florian Bruckner[1,4], Dieter Suess[1,4], Andrii V. Chumak[1] & Claas Abert[1,4] ✉

The inverse design approach in magnonics exploits the wave nature of magnons and machine learning to develop logical devices with functionalities that exceed the capabilities of analytical methods. While promising for analog, Boolean, and neuromorphic computing, current implementations face memory limitations that hinder the design of complex systems. This study presents a level-set parameterization method for topology optimization, combined with an adjoint-state approach for memory-efficient simulation of magnetization dynamics. The framework is implemented in `NeuralMag`, a GPU-accelerated micromagnetic solver featuring a nodal finite-difference scheme and automatic differentiation tools. To validate the method, we optimized the shape of a magnetic nanoparticle by applying constraints to the objective function, and designed a 300 nm-wide yttrium iron garnet demultiplexer achieving frequency-selective spin-wave separation. These results highlight the algorithm's efficiency in exploring local minima across various initial configurations, establishing its utility as a versatile tool for the inverse design of magnonic logic devices.

The use of coherent magnetization oscillation phenomena, known as spin waves, as a tool for writing, receiving, and transmitting information has been extensively studied[1–5]. This research is driven by the search for viable frameworks for beyond-CMOS technology. Initial efforts have evolved from millimeter-scale devices that mimic conventional logic[6,7] to micrometer-scale processing units[8–11]. However, further miniaturization to nanometers introduces additional challenges and complexities in the design of such devices. Additionally, the integration of features such as non-linearity and multi-frequency operation adds to this complexity, making it difficult to create geometries that can leverage these phenomena simultaneously using conventional design approaches.

In this context, the inverse design approach has emerged as a promising solution for creating optimized geometries. This method involves two primary steps: parameterizing the design region and optimizing these parameters through iterative feedback calculations based on an objective function that defines the device's required functionality[12].

The inverse design method has already demonstrated its efficiency in photonics, where it has been used to create on-chip devices for tasks such as mode multiplexing[13], angular wave separation[14,15], and light polarization[16]. The method's versatility has been further proven through its applications in various other areas[12]. More recently, this approach has been adapted to magnonics, where it utilizes the similarities with integrated photonics to tackle challenges such as spin wave property-selective guiding[17] and vowel recognition in both linear and non-linear regimes[18–20].

However, existing inverse design algorithms in magnonics face significant hardware limitations[18], as they require saving all computational steps, resulting in substantial memory usage. These constraints hinder the exploration of more complex geometries and limit the size of the simulation system. To address these challenges, a conceptually new algorithm is needed–one that increases computational depth while minimizing hardware demands, thereby overcoming the limitations of current optimization approaches. This new algorithm should also be capable of solving a wide range of inverse-design problems, including shape and topology optimization, which may demand comprehensive micromagnetic simulations.

To achieve this, integrating general-purpose micromagnetic software with an optimization technique capable of handling geometry optimization tasks is a promising direction. Such tasks often involve the homogeneous movement of boundaries between regions and are commonly studied in disciplines such as fluid mechanics, material science, meteorology, and computer vision[21]. In this regard, the level-set method, originally introduced for studying flame propagation[22], is particularly relevant. This method describes the evolution of a higher-dimensional level-set function (LSF) that implicitly defines the interface boundary through its iso-contour. One of its main advantages over explicit methods is its ability to handle topological changes seamlessly, such as boundary merging, the formation of new shapes, or the disappearance of existing ones[23].

[1]Faculty of Physics, University of Vienna, 1090 Vienna, Austria. [2]Vienna Doctoral School in Physics, University of Vienna, 1090 Vienna, Austria. [3]Faculty of Sciences, University of Oviedo, 33003 Oviedo, Spain. [4]Research Platform MMM Mathematics - Magnetism - Materials, University of Vienna, Vienna, Austria.
✉e-mail: andrey.voronov@univie.ac.at; claas.abert@univie.ac.at

Traditionally, the optimization of the LSF $\Phi$ is guided by the Hamilton-Jacobi equation: $\Phi_t + \nabla_x \Phi \cdot \mathbf{v} = 0$, where $\mathbf{v}$ represents the velocity field of the boundaries. For numerical implementation, the gradient descent method is a straightforward and effective approach to iteratively update the LSF[24]. This approach involves calculating the gradient of the objective function $J$, which quantifies the difference between the target design solution and the simulated one.

Applying the current optimization method to micromagnetic simulations presents challenges when solving the Landau-Lifshitz-Gilbert (LLG) equation, which governs magnetization dynamics and forms the foundation of micromagnetic software[25]. Calculating gradients with respect to the LLG solution, a partial differential equation (PDE) in space and time, significantly increases memory usage, complicating the optimization of complex geometries. To address this, the adjoint-state method has been chosen as an effective technique for gradient calculation[26]. This method, which has seen applications in fields such as fluid mechanics[27,28], aerodynamics[29], seismology[30], and geophysics[31], computes gradients by solving a second adjoint PDE backward in time rather than backpropagating through all LLG solver operations. Consequently, it enables optimization with a constant memory footprint as a function of simulation time and controllable calculation error[32].

The present study provides an algorithm that combines the level-set and adjoint methods for efficiently solving inverse design problems using micromagnetics. It is built upon the GPU-accelerated micromagnetic software `NeuralMag`[33], which makes use of automatic differentiation for gradient computation provided by PyTorch backend. `NeuralMag`'s nodal finite-difference discretization scheme facilitates efficient boundary treatment between different materials, a feature heavily employed in this research.

This algorithm was tested on two optimization tasks designed to demonstrate the level-set method's efficiency in topology variation. The first task focused on optimizing the hysteresis curve by adjusting the shape of a nanoparticle, revealing the algorithm's smooth convergence toward a local minimum. The second task, involving spin-wave propagation, introduced additional complexity and highlighted the benefits of the adjoint-state method.

The results validate the proposed optimization software, establishing it as a versatile and universal tool for future inverse-design studies in magnonics.

## Results
### Level-set method
The level-set method is a versatile tool used to track and model the boundaries between different regions or materials. It is especially useful because it can easily handle changes in shape, such as merging or splitting, which are common in many scientific fields like fluid dynamics and material science. This method represents the boundary as a contour of a higher-dimensional function, allowing for smooth and continuous changes[23,34].

The main principle of the level-set method is that the boundary $\Gamma$ between two regions (or materials) $A$ and $B$, the position of which is to be optimised during the simulation, is defined as the "zero-level-set" of the LSF $\Phi(x,y)$ Fig. 1(a):

$$\begin{cases} \Phi(x,y) < 0 & \text{if } (x,y) \in A, \\ \Phi(x,y) = 0 & \text{if } (x,y) \in \Gamma, \\ \Phi(x,y) > 0 & \text{if } (x,y) \in B, \end{cases} \quad (1)$$

where the LSF $\Phi(x,y)$ is a global higher-dimensional function that covers the whole design region. This study focuses on the optimization of 2D topologies for thin-film logic devices. However, the level-set method can be efficiently extended to 3D applications, as demonstrated in previous studies[35].

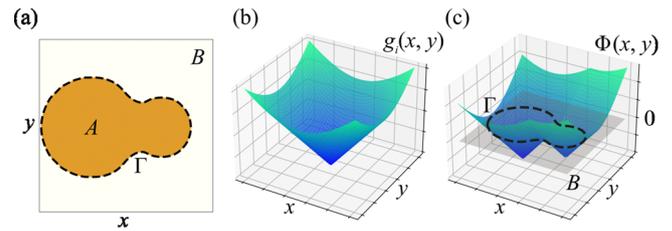

Fig. 1 | Construction of the level-set function. a Representation of material domains $A$ and $B$ with an interface $\Gamma$ optimized during the simulation. b Visualization of a 3D radial basis function $g_i(x,y)$. c Global 3D level-set function $\Phi(x,y)$ constructed from RBFs, with the LSF sliced at the "zero-level-set" to define the interface between material domains.

One commonly used and effective method for representing the LSF is through radial basis functions (RBFs). This approximation allows for the global smoothness of the LSF and proves to provide a better convergence of the subsequent optimisation process[36]. RBFs are radially symmetric functions centred at a certain point on the mesh. Different functions can be utilised as RBFs, but the multiquadratic spline appears to be the most promising choice[37]. The following formula defines the RBF used in this work:

$$g_i(x,y) = s_i - \sqrt{(x-x_i)^2 + (y-y_i)^2 + c^2}, \quad (2)$$

where $x_i$ and $y_i$ are the coordinates of the RBF center, $c$ is a constant, and $s_i$ is the amplitude of the $i$-th RBF, which is optimized during the simulation. This RBF has a cone-like shape (Fig. 1b), and adjusting $s_i$ alters the cone's height, modifying the LSF and the resulting topology.

Typically, the LSF is expressed as the sum of all RBFs: $\Phi(x,y) = \sum_i^n g_i(x,y)$. However, this global approach limits local topology adjustments, as changes to a single RBF impact the entire LSF. To address this, the LSF can be defined as the maximum value of all RBFs at a given point: $\Phi(x,y) = \max_i g_i(x,y)$. However, to subsequently perform the gradient calculation one has to use the differentiable operation to calculate the LSF. In the present research, the $p$-norm is used to approximate the maximum value of all RBFs at every point[38]:

$$\Phi(x,y) = \left[\sum_{i=1}^n (g_i(x,y) + \Delta\phi)^p\right]^{1/p} - \Delta\phi, \quad (3)$$

where $\Delta\phi$ is the artificial shift used to ensure the positive value inside the square brackets of the equation. Since this shift introduces an additional error, $\Delta\phi$ is chosen as the absolute value of the smallest RBF in the design region. Moreover, a value of $p = 90$ is used in the present work to calculate the norm. The LSF values are kept within the range of $-1$ to $1$ through renormalization after each optimization step to control boundary movement speed and avoid issues related to overly steep or flat LSFs[23].

After constructing the LSF, the next step is to map it to a concrete 2D topology for micromagnetic simulations. One option of achieving this is to use the Heaviside function that allows for direct mapping of the Eq. (1) and a sharp boundary between the materials. However, since this operation is non-differentiable one has to use smooth functions that, on one hand, introduce a non-zero width of the boundary but, on the other hand, open the way for applying the gradient-type optimisation techniques. In the present study, a smooth sigmoid activation function is used to map the LSF on the design region:

$$\psi(x,y) = \frac{1}{1 + \exp(-a\,\Phi(x,y))}, \quad (4)$$

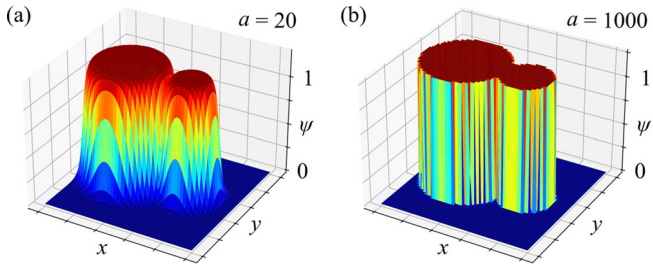

**Fig. 2 | Mapping procedure of the LSF function $\Phi(x, y)$ onto the design region to determine the boundary position. a** Smooth boundary between material domains ($\psi = 0$ or 1) with $a = 20$ in Eq. (4). **b** Sharp boundary using $a = 1000$ or the Heaviside function.

where the additional coefficient $a$ is used to control the smoothness of the boundary (Fig. 2) and, thus, the convergence rate of the topology optimisation. In this work, $a = 50$ is used. Then the material parameters (e.g. saturation magnetization $M_s$, exchange constant $A_{ex}$) are multiplied by the resulting mapping $\psi(x, y)$ in the micromagnetic simulation, determining different material domains. Consequently, the optimisation procedure adjusts the amplitudes **s** of different RBFs governing the shape of the global LSF and the boundaries $\Gamma$ between materials.

### Optimisation with adjoint method

The amplitudes **s** of the RBFs, and consequently the boundary $\Gamma$ between domains, are updated using the gradient descent method. The first step in this process is defining the objective function $J$, which describes the behavior of the desired design relative to the current system output. The objective function quantifies the loss of the structure compared to its required functionality, thereby guiding the optimization procedure. For each specific task, users must construct a tailored objective function that governs the optimization process.

The parameters **s** are updated iteratively following a forward simulation, according to the gradient descent method[39]:

$$\mathbf{s}^{j+1} = \mathbf{s}^j - \alpha \frac{\partial J}{\partial \mathbf{s}}\bigg|_{\mathbf{s}=\mathbf{s}^j}, \quad (5)$$

where index $j$ indicates the optimisation step, $\alpha$ is the learning rate that controls the convergence of the process. The efficient calculation of the gradient is not a trivial but nevertheless crucial task that strongly affects the performance of the simulation software[40].

This study employs the adjoint-state method to solve the LLG equation efficiently and compute the gradients relative to the optimisation parameters. It is a widely used technique for efficiently computing gradients in optimization problems, especially those involving partial differential equations (PDEs)[26,31,41]. It is particularly advantageous in scenarios where direct gradient computation would be computationally prohibitive due to the high dimensionality of the problem.

The method works by solving the system of adjoint equations:

$$\begin{cases} \frac{\partial \mathbf{m}}{\partial t} = \mathcal{L}(t, \mathbf{m}, \mathbf{s}), \\ \frac{\partial \mathbf{a}}{\partial t} = -\mathbf{a}^T \cdot \frac{\partial \mathcal{L}(t, \mathbf{m}, \mathbf{s})}{\partial \mathbf{m}}, \end{cases} \quad (6)$$

where the first equation represents the LLG equation for the unit magnetization **m**, solved in the time domain from $t = t_0$ to $t = T$. The adjoint parameter $\mathbf{a}(t) = \partial J / \partial \mathbf{m}(t)$ is critical for gradient computation and its dynamics is governed by the second equation in the system Eq. (6). Its value at $t = t_0$ has to be computed by the second call of the solver, which should run backwards in time with the initial condition $\mathbf{a}(T) = \partial J/\partial \mathbf{m}(T)$. This approach eliminates the need to store intermediate values of $\mathbf{m}(t)$ in memory, as they can be recomputed during the backward pass of the adjoint equation.

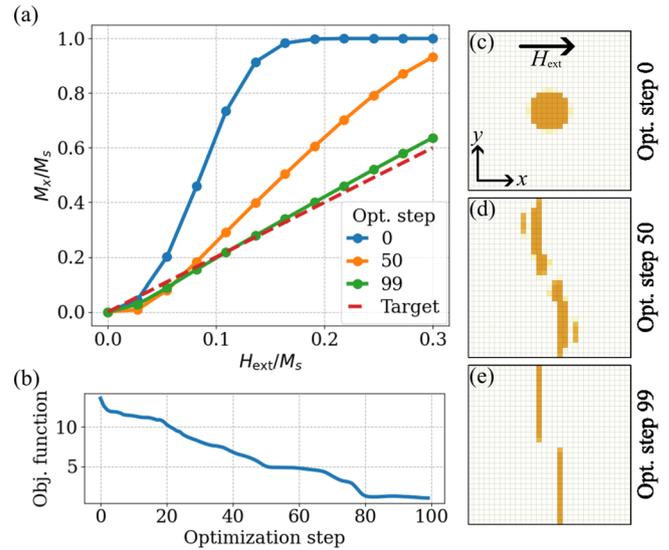

**Fig. 3 | Optimization process of the magnetic particle's topology based on the magnetization saturation curve. a** Evolution of the hysteresis curve across different optimization steps, compared with the target curve. **b** Objective function progression over 100 optimization steps. (**c**–**e**) Evolution of the particle geometry and the system configuration during optimization.

Gradients of the function with respect to the parameter **s** are then calculated using the following integral[32,33]:

$$\frac{\partial J}{\partial \mathbf{s}} = -\int_T^{t_0} \mathbf{a}(t)^T \frac{\partial \mathcal{L}(t, \mathbf{m}, \mathbf{s})}{\partial \mathbf{s}} dt, \quad (7)$$

which is substituted into Eq. (5) to perform the optimization step. Further discretization and numerical evaluation of the integral in Eq. (7) are performed using the `torchdiffeq` package[32]. This approach replaces the conventional chain rule, which would require significant memory to compute gradients for complex designs. The adjoint method thus ensures efficient memory use and better control over computational errors during simulations.

The integration of the level-set and adjoint methods provides a robust framework for performing micromagnetic simulations for various topology optimization tasks, regardless of their complexity.

### Stoner-Wohlfarth particle

The first validation task for the proposed software is optimizing the hysteresis curve of a small magnetic particle. The saturation curve of such particles is strongly influenced by shape anisotropy (therefore, by the particle's topology) and the direction of the applied external field, as described by the Stoner-Wohlfarth model[42]. In this study, we investigate the longitudinal hysteresis, measuring magnetization $M$ along the direction of the external magnetic field $H_{ext}$. According to the model, in this configuration, the hysteresis loop collapses into a single line when the magnetic field is applied along the ellipsoidal particle's shorter axis (so-called hard axis loop), with the magnetization curve following the linear law $M(H_{ext}) = bH_{ext}$[43]. Here, the constant $b$ depends on the particle's ellipticity and takes the value $b = 2$ in the limiting case of an infinitely long magnetic wire.

This limiting case is selected as the target for the optimization software. Thus, the objective of this task is to design a particle shape that produces a hysteresis curve following the law $M^{target}(H_{ext}) = 2H_{ext}$, shown as the target curve in Fig. 3(a). Given that the final particle shape is predictable through theoretical models, this topology optimization serves as a reliable validation point for the inverse-design software.

The geometry under investigation is shown in Fig. 3(c). The design region mesh consists of a $30 \times 30 \times 1$ cubic cells with a 3 nm size of the individual cell. The magnetic particle, initially circular, is surrounded by non-magnetic air, and Permalloy (NiFe) is selected as the magnetic material with saturation magnetization $M_s = 800$ kA/m and exchange constant $A_{ex} = 13$ pJ/m. The external magnetic field $H_{ext}$ is applied along the x-axis with the particle's average magnetization component $M_x$ plotted in Fig. 3(a).

For optimization, the LSF was initialized over the entire mesh using $20 \times 20$ RBFs arranged in an evenly spaced grid. The objective function $J$ was defined to measure the difference between the particle's magnetization curve $M_x(H_{ext})$ and the target function $M^{target}(H_{ext})$:

$$J = \sum_{j=0}^{N} \frac{|M_x(H_{ext}^j) - M^{target}(H_{ext}^j)|}{M^{target}(H_{ext}^j)N}, \quad (8)$$

where in each epoch the summation is performed over $N$ simulation steps, with the external field ramping from $H_{ext}^{j=0} = 0$ to $H_{ext}^{j=N} = 0.3M_s$. With this setup, the software achieved a stable objective function value $J$ after 100 iterations, as shown in Fig. 3b. The design optimization process (Fig. 3c–d) concludes with two elongated magnetic particles resembling wires. Since the algorithm primarily finds local minima, this solution represents one of multiple possible geometries that can replicate the target hysteresis curve $M^{target}(H_{ext})$.

These results demonstrate the effectiveness of the algorithm in solving the magnetization optimization problem. Additionally, the intermediate design in Fig. 3d highlights one of the level-set method's unique features: topology separation, where a single unified shape can evolve into multiple distinct particles to meet the objective function. Utilizing such features expands the design space significantly, allowing for a range of viable non-unique solutions in optimization tasks.

Alternatively, if one aims to restrict the solution space and achieve a specific design, regularization techniques can be employed. In the present example, the optimal geometry satisfying the objective function would be a uniform vertical wire along the axis of the design region. Narrowing the design space is accomplished by introducing additional constraints to the objective function (Eq. (8)) or by dynamically adjusting the learning rate $\alpha$ during the optimization process to facilitate smoother convergence.

To illustrate the impact of regularization on the final solution, the particle shape optimization task was repeated using a modified objective function $J$. In this version, the baseline term measuring the difference between the particle's and the target magnetization curves (Eq. (8)) was augmented with additional constraints for size and center of mass. The size constraint restricts the nanoparticle's surface area to match that of a 1-cell-width vertical wire along the simulation mesh, while the center constraint ensures the particle's center of mass aligns with the center of the design region:

$$J = \zeta \underbrace{\sum_{j=0}^{N} \frac{|M_x(H_{ext}^j) - M^{target}(H_{ext}^j)|}{M^{target}(H_{ext}^j)N}}_{\text{Curve difference}} + \underbrace{\xi(S - S^{target})^2}_{\text{Size constraint}} \\ + \underbrace{\nu[|x_c - x_0| + |y_c - y_0|]}_{\text{Center constraint}}. \quad (9)$$

Here, $S$ and $\{x_c, y_c\}$ represent the surface area and center of mass coordinates of the magnetic particle, while $S^{target}$ and $\{x_0, y_0\}$ are the target values for these parameters. The weights $\zeta$, $\xi$, and $\nu$ determine the relative influence of each constraint on the total objective function value (Fig. 4a).

Additionally, the learning rate $\alpha$ was set to decrease after every ten optimization steps to ensure smooth convergence of the algorithm to the minimum value (Fig. 4b).

Applying these regularization techniques to the particle optimization resulted in the desired uniform long wire along the design region (Fig. 4c). Although the optimization process required 250 steps to converge to the minimum–due to the increased stiffness of the objective function introduced by the constraints–the algorithm successfully reached the desired solution by gradually adjusting the learning rate at each step.

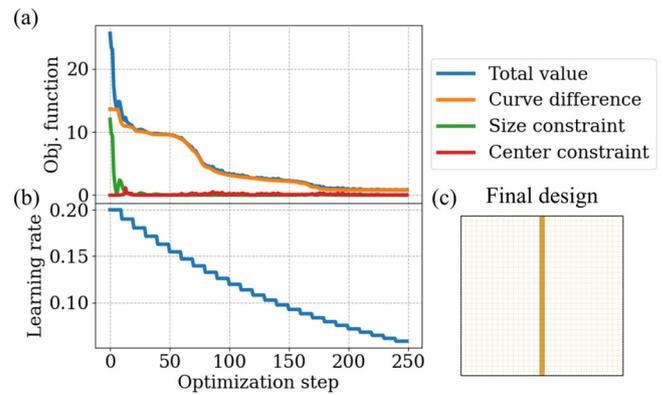

**Fig. 4 | Impact of modified constraints on the particle topology optimization process. a** Contribution of the additional constraints to the total objective function value during optimization. **b** Evolution of the learning rate $\alpha$ throughout the optimization process. **c** Final design of the nanoparticle obtained using the decreasing learning rate technique and the modified objective function (Eq. (9)) with the following weights of the constraints: $\zeta = 10$, $\xi = 0.05$, $\nu = 1$.

The inclusion of constraints provides users with an additional level of control, enabling them to refine the solution space and selectively identify designs that offer the best functionality for specific applications.

### Demultiplexer example

Next, a more complex example involving coherent spin-wave propagation demonstrates the capability of the proposed software to design nanoscale logic devices. This task focuses on frequency-dependent spin-wave separation, a problem first solved via inverse design by Qi Wang et al.[17] using a binary search algorithm as the optimization tool.

In this example, two spin waves with frequencies $f_1 = 2.6$ GHz and $f_2 = 2.8$ GHz are excited simultaneously in a 300 nm-wide, 100 nm-thick conduit made of yttrium iron garnet ($Y_3Fe_5O_{12}$, $M_s = 140$ kA/m, $A_{ex} = 3.5$ pJ/m), guiding them into the design region (Fig. 5(a)). This design region, measuring $1 \times 1$ $\mu m^2$, is optimized to have a specific distribution of air holes that create an interference pattern directing spin waves to different output conduits (labeled as "top" and "bottom" in Fig. 5(a)) based on their frequency, effectively performing the function of a demultiplexer. To enable forward volume geometry and achieve isotropic spin wave propagation, an external field of 200 mT is applied perpendicularly to the surface in this study. The simulation uses a total mesh size of $512 \times 64 \times 1$ with each cell measuring $20 \times 20 \times 100$ nm$^3$.

To define the design region, the LSF was constructed using $20 \times 20$ RBFs arranged in an evenly spaced array. The initial RBF amplitudes **s** were set to form a $4 \times 4$ grid of holes within the design region (Fig. 5b, c).

The excitation signal applied to the input conduit was kept at a low field strength (in the sub-mT range) to ensure that spin waves remained within the linear regime, with magnetization oscillation angles below 1 degree. This approach minimizes non-linear effects that could interfere with the desired device functionality[44]. In this study, the frequency-selective guiding of spin waves is achieved solely through wave-front interference and diffraction caused by the air holes generated in the design region by the LSF.

The objective function $J$, constructed to achieve the demultiplexer functionality in the gradient-based optimization algorithm, is defined as follows:

$$J = \left(A_{TO}^{f_1} - A_{BO}^{f_1}\right) + \left(A_{BO}^{f_2} - A_{TO}^{f_2}\right), \quad (10)$$

where $A$ represents the sum of the magnetization oscillation FFT values within a specified frequency window (around $f_1$ or $f_2$) calculated at the top or bottom outputs (TO or BO). These frequency windows are illustrated in Fig. 6(a, b). The objective function is designed to maximize the amplitude of the spin wave at frequency $f_1$ in the bottom output and at frequency $f_2$ in the top output, while minimizing cross-talk between the two output conduits.

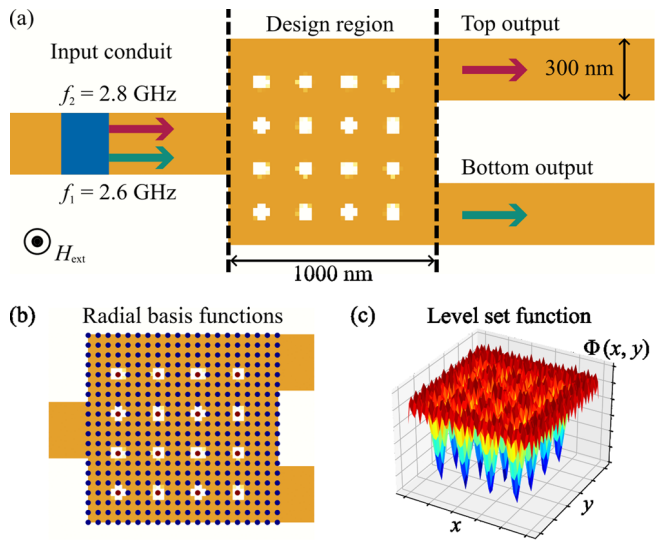

**Fig. 5 | Visualization of the simulation domain and level-set function configuration for spin-wave optimization. a** Schematic representation of the simulation domain, showing: the excitation of two spin waves in the blue region of the input conduit; the design region with the initial hole distribution used for optimization; and the two output conduits for frequency-based spin wave detection. **b** Distribution of the centers of all 400 RBFs used to define the LSF within the design region, with dot colors representing the amplitude values $s_i$ of the corresponding $i$-th RBF. **c** The LSF defining the magnetic topology.

Using the initial design shown in Fig. 5(a) and the objective function in Eq. (10), the optimization algorithm reached a stable local minimum within 100 optimization steps (Fig. 6(a)). This optimized state of the LSF produced the geometry shown in the inset of Fig. 6(a). The magnetization FFT values at different outputs (Fig. 6c, d) reveal clear frequency-based spin-wave separation, with amplitudes differing by an order of magnitude between the target frequencies. Figure 6(b) additionally shows the color maps of the spin-wave propagation through the optimized design region for different excitation frequencies $f_1$ and $f_2$.

For optimization, the objective function in Eq. (10) accounted not only for the FFT value at a single frequency but for the sum of FFT values within windows around $f_1$ and $f_2$. These windows are illustrated with colored regions in Fig. 6f, g with the green corresponding to the window around $f_1$ = 2.6 GHz and red to the one around $f_2$ = 2.8 GHz. This approach suppresses neighboring frequencies during optimization, enhancing the robustness of the final design.

This robustness was further validated through frequency-sweep simulations (Fig. 6e), which demonstrated that the spin-wave separation is maintained even with slight deviations from the target frequencies, creating an operating window of approximately 70 MHz. This tolerance is critical for practical applications, where slight deviations are common in nanoscale fabrication.

Additionally, the algorithm displayed strong convergence properties across different initial designs. Figure 7 shows the optimization process for the demultiplexer functionality with two distinct initial LSF configurations: one with a 2 × 2 array of holes and another with a single central hole. In both cases, the optimization successfully found a design that met the objective function requirements. Remarkably, even starting with a single hole, the algorithm efficiently introduced the minimal number of new elements required to achieve the desired functionality Fig. 7(d, h).

These results highlight an important property of the proposed algorithm: while selecting an appropriate initial geometry can expedite optimization, the level-set method's inherent flexibility enables various initial configurations to converge to successful outcomes, thereby enhancing the robustness of the approach. This phenomenon arises from the nonlinear dependence of the objective function $J$ on the RBF amplitudes **s**. The nonlinearity creates a complex landscape for $J$, with multiple local minima that are explored by the gradient descent method. The interplay of these factors explains the observed effect: optimizations starting from different initial designs converge to distinct local minima. While the structure resulting from the 4 × 4 design optimization demonstrates the best spin-wave separation performance, each final design satisfies the objective function.

Additional benefit is that the gradient descent approach facilitates the simultaneous optimization of the entire design region at each step, requiring significantly fewer simulations to achieve convergence compared to the direct binary search method previously used[17]. Furthermore, the resulting designs exhibit smoother, more rounded features, which are better suited for fabrication using lithography techniques.

## Discussion

This study presents a novel algorithm that combines the NeuralMag micromagnetic solver with the comprehensive topology optimization capabilities of the level-set method. This approach significantly increases the degrees of freedom in shape optimization for complex structures and single particles by parameterizing the design region and constructing a global level-set function, which is then evolved through gradient-based optimization.

The level-set method provides a straightforward means of handling event-type phenomena that typically challenge conventional shape optimization methods, such as hole nucleation and topology merging, which can create discontinuities at material interfaces. These phenomena are intrinsically managed by the algorithm's design, enhancing its versatility.

To demonstrate the algorithm's robustness, we applied it to two distinct optimization tasks. The first task involved optimizing the shape of a single particle based on its magnetization saturation curve. This example illustrates how the algorithm can achieve different local minima that satisfy the objective function, facilitated by the algorithm's intrinsic handling of event-type topology changes.

The second example focused on designing a micrometer-sized demultiplexer that directs spin waves according to their frequency. In this case, the holes in the structure serve as diffraction centers, with their optimized positions precisely controlling the device's spin-wave guiding properties. This example highlights the algorithm's capacity for addressing complex optimization tasks involving spin-wave propagation and the design of functional logic devices.

The ability to handle complex magnetic tasks is enabled by the adjoint-state method for solving the Landau-Lifshitz-Gilbert equation, allowing efficient backpropagation and gradient calculation without overwhelming computational resources.

The features demonstrated in these magnetic problems highlight the significant potential of the proposed algorithm for future applications in inverse design. One promising direction is the extension of the level-set method to 3D structures[34], which would enable the optimization of complex heterostructures. This is particularly relevant for systems influenced by the Dzyaloshinskii-Moriya interaction, which plays an important role in guiding topology optimization in hysteresis loop measurements[45]. Notably, DMI is already implemented in the NeuralMag software used in this study, making it possible to incorporate this interaction into future inverse-design optimizations seamlessly. The ability to optimize such geometries could be highly beneficial for designing advanced materials and devices with intricate magnetic structures, further broadening the applicability of the method.

Another notable feature of the proposed solver is its capability to handle not only PDEs but also stochastic differential equations through a modified adjoint method[46,47]. This extension allows for the implementation of finite-temperature optimization tasks, where thermal fluctuations are incorporated into the LLG equation as an additional effective field, computed via the Langevin equation[48–50]. An alternative approach to account for temperature effects in micromagnetics involves using temperature-dependent material parameters, such as the saturation magnetization and damping constant, which can be extracted from experiments or thermodynamic models. By integrating stochastic thermal fields or temperature-dependent material properties, the optimization process can more accurately reflect experimental conditions and expand the range of problems that can be addressed using inverse design.

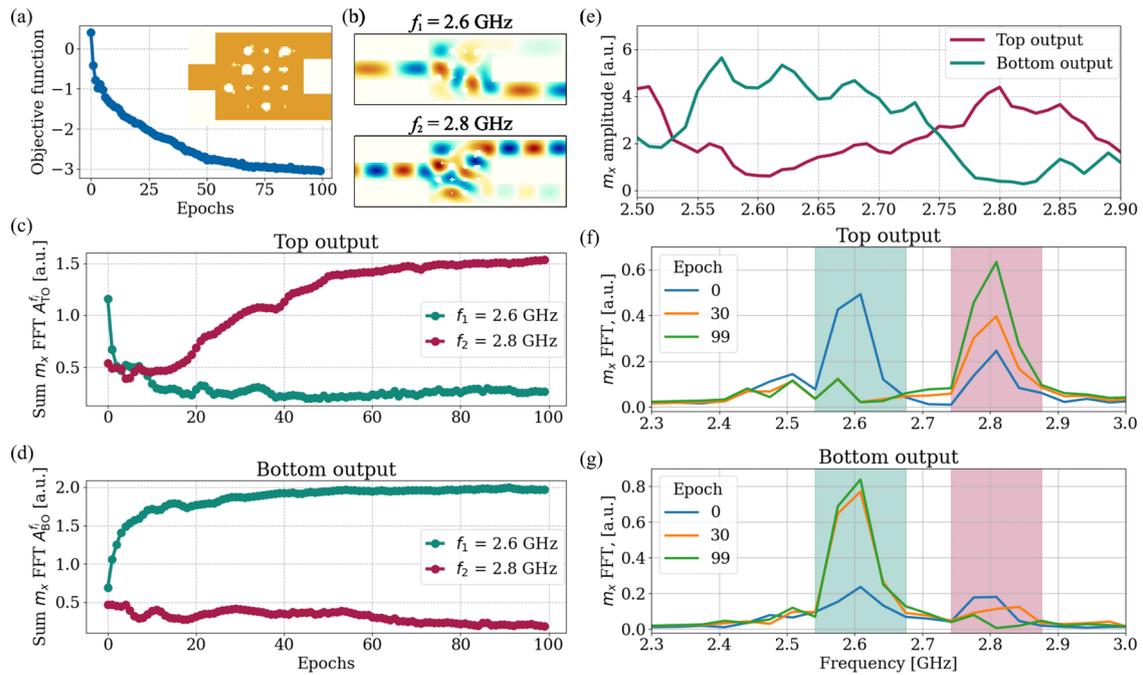

**Fig. 6 | Demultiplexer optimization process starting with a 4 × 4 array of holes. a** Objective function evolution over 100 optimization steps with a final optimized design produced by the algorithm as an inset. **b** Spin-wave propagation map of the optimized design for both excitation frequencies $f_1$ and $f_2$. Spin-wave amplitudes at the top (**c**) and bottom (**d**) outputs over the optimization process. **e** Frequency-dependent magnetization oscillation amplitude at the top and bottom outputs. FFT spectra of the magnetization at the top (**f**) and bottom (**g**) outputs for three different optimization steps, showing frequency windows used for objective function calculations (green for $f_1$ and red for $f_2$).

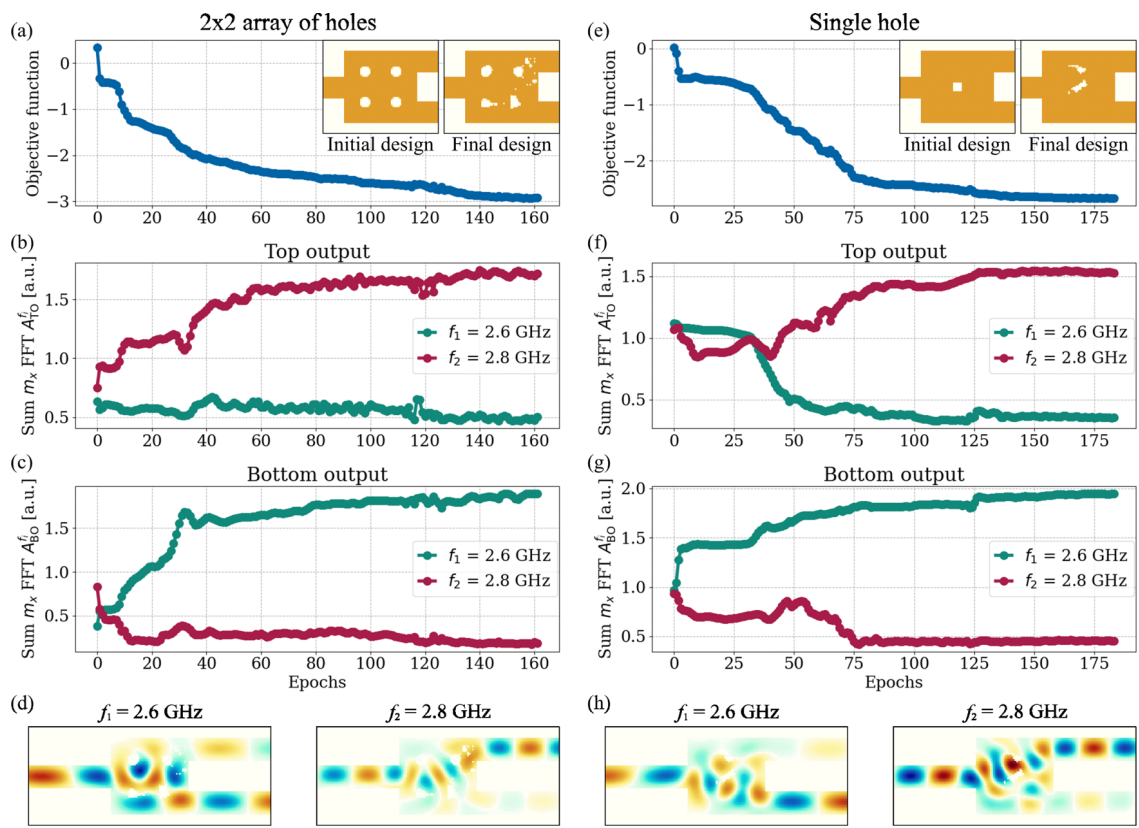

**Fig. 7 | Demultiplexer optimization starting from two different initial designs: (left panel) a 2 × 2 array of holes and (right panel) a single square hole at the center of the design region. (a/e)** Objective function evolution, with insets showing the initial and final designs of the demultiplexer. (**b, c/f, g**) Spin-wave separation during the optimization process. (**d/h**) Spin-wave propagation maps of the final designs for the two excitation frequencies under study.

Furthermore, the utilization of nonlinear spin-wave phenomena offers new possibilities for developing complex magnonic logic gates, such as nonlinear switches and half-adders[17]. These advancements could pave the way for more sophisticated and functional magnonic devices with arbitrary complexity. By expanding the scope of inverse design, the proposed framework holds promise for applications beyond traditional logic, including neuromorphic computing and reconfigurable magnonic circuits, ultimately contributing to the development of next-generation spintronic technologies.

## Methods
### Micromagnetic simulations
All results presented in this work were obtained through micromagnetic simulations using the GPU-accelerated software package `NeuralMag`[33]. The magnetization dynamics were computed by numerically solving the Landau-Lifshitz-Gilbert (LLG) equation:

$$\frac{\partial \mathbf{m}}{\partial t} = -\gamma\, \mathbf{m} \times \mathbf{H}_{\text{eff}} + \alpha\, \mathbf{m} \times \frac{\partial \mathbf{m}}{\partial t}, \qquad (11)$$

where $\mathbf{m}$ is the unit magnetization vector, $\gamma$ is the gyromagnetic ratio, $\alpha$ is the Gilbert damping constant, and $\mathbf{H}_{\text{eff}}$ is the effective magnetic field. The effective field for the tasks under consideration includes contributions from the Zeeman interaction, the demagnetizing (dipole-dipole) field, and the exchange interaction.

The numerical solution of the LLG equation was performed using a nodal finite-difference discretization scheme and an adaptive step-size time integration method to ensure both numerical stability and efficiency. The spatial discretization was adapted to the geometry and resolution requirements of each simulation scenario.

For the optimization procedure, gradients of the objective function with respect to the optimized parameters were computed using the adjoint-state method in combination with PyTorch's automatic differentiation capabilities. The parameter update was performed using the Adam optimizer, a stochastic gradient descent algorithm with adaptive moment estimation[51].

## Data availability
No datasets were generated or analysed during the current study.

## Code availability
The underlying code for this study is not publicly available but may be made available to qualified researchers on reasonable request from the corresponding author.

## Acknowledgements
This research was funded in whole or in part by the Austrian Science Fund (FWF) projects: 10.55776/P34671, 10.55776/I6068, MagFunc [10.55776/I4917] and IMECS [10.55776/PAT3864023]. The computational results presented were achieved using the Vienna Scientific Cluster (VSC-5). For the purpose of open access, the author has applied a CC BY public copyright license to any Author Accepted Manuscript version arising from this submission.

## Author contributions
AAV conducted all numerical simulations presented in the paper, implemented the level-set approach, and developed the framework for the demultiplexer simulations. MCS created the framework for the Stoner-Wohlfarth nanoparticle topology optimization task. FB contributed to interpreting the simulation results. CA developed the NeuralMag software used for micromagnetic simulations and provided theoretical support. DS and AVC identified a scientific challenge, while CA led the project. All authors contributed to the scientific discussion and commented on the manuscript.

## Competing interests
The authors declare no competing interests.

## Additional information
**Correspondence** and requests for materials should be addressed to Andrey A. Voronov or Claas Abert.

**Reprints and permissions information** is available at http://www.nature.com/reprints

**Publisher's note** Springer Nature remains neutral with regard to jurisdictional claims in published maps and institutional affiliations.